\documentclass[fleqn,usenatbib]{mnras}

\usepackage[T1]{fontenc}
\usepackage{ae,aecompl}
\usepackage{graphicx}
\usepackage{epstopdf}
\usepackage{amsmath}
\usepackage{amssymb}
\usepackage{enumerate}
\usepackage{xcolor}
\usepackage[normalem]{ulem}
\title[Late-time Afterglow Evolution]{Late-time Evolution of Afterglows from Off-Axis Neutron-Star Mergers}
\author[G.P. Lamb, I. Mandel \& L. Resmi]{Gavin P Lamb$^{1}$\thanks{E-mail: gpl6@leicester.ac.uk}, Ilya Mandel$^{2, 3}$, and Lekshmi Resmi$^{4}$
\\
$^{1}$Department of Physics and Astronomy, University of Leicester, University Road, Leicester, LE1 7RH, UK
\\
$^{2}$Institute of Gravitational Wave Astronomy and School of Physics and Astronomy, University of Birmingham,\\ Birmingham, B15 2TT, UK
\\
$^{3}$Monash Centre for Astrophysics, School of Physics and Astronomy, Monash University, Clayton, Victoria 3800, Australia
\\
$^{4}$Indian Institute of Space Science \& Technology, Trivandrum 695547, India
}

\date{Accepted XXX. Received YYY; in original form ZZZ}
\pubyear{2018}
\begin{document}
\label{firstpage}
\pagerange{\pageref{firstpage}--\pageref{lastpage}}
\maketitle

\begin{abstract}
Gravitational-wave detected neutron star mergers provide an opportunity to investigate short gamma-ray burst (GRB) jet afterglows without the GRB trigger.
Here we show that the post-peak afterglow decline can distinguish between an initially ultra-relativistic jet viewed off-axis and a mildly relativistic wide-angle outflow.
Post-peak the afterglow flux will decline as $F_\nu \propto t^{-\alpha}$.
The steepest decline for a jet afterglow is $\alpha>3p/4$ or $> (3p+1)/4$, for an observation frequency below and above the cooling frequency, respectively, where $p$ is the power-law index of the electron energy distribution.
The steepest decline for a mildly relativistic outflow, with initial Lorentz factor $\Gamma_0\lesssim 2$, is $\alpha\lesssim(15p-19)/10$ or $\alpha\lesssim(15p-18)/10$, in the respective spectral regimes.
If the afterglow from GW170817 fades with a maximum index $\alpha > 1.5$ then we are observing the core of an initially ultra-relativistic jet viewed off the central axis, while a decline with $\alpha\lesssim 1.4$ after $\sim 5$--10 peak times indicates that a wide angled and initially $\Gamma_0\lesssim 2$ outflow is responsible. 
At twice the peak time, the two outflow models fall on opposite sides of $\alpha \approx 1$. So far, two post-peak X-ray data points at 160 and 260 days suggest a decline consistent with an off-axis jet afterglow.  
Follow-up observations over the next 1--2 years will test this model.
\end{abstract}

%%%%%%%%%%%%%%%%%%%%%%%%%%%%%%%%%%%%%%%%%%%%%%%%%%%%%%%%%%%%%%%%%%%%%%%%%%%
\section{Introduction}

Binary neutron star or neutron star -- black hole mergers are the likely progenitors systems for gamma-ray bursts (GRBs) with a duration $\lesssim 2$ s, the so-called short GRBs \citep{eichler1989, narayan1992, mochkovitch1993, bogomazov2007, nakar2007}.  GRB afterglows have typically been detected following a GRB triggered search \citep{berger2014} \citep[however, see][ where the afterglow was detected independent of a high energy trigger]{cenko2013, cenko2015}.
Gravitational-wave (GW) detected mergers involving a neutron star provide a new trigger for afterglow searches free of the inclination constraints of the GRB trigger, where the GRB is highly beamed from within an ultra-relativistic jet.

The GW detected merger of a binary neutron star system, GW170817 \citep{abbott2017}, was associated with a weak GRB \citep{abbott2017a, fermi2017, savchenko2017}, an optical afterglow over the days following the merger \citep[e.g.][]{arcavi2017, buckley2018, chornock2017, coulter2017, covino2017, cowperthwaite2017, drout2017, diaz2017, evans2017, kilpatrick2017, lipunov2017, mccully2017, nicholl2017, pian2017, shappee2017, smartt2017, soares-santos2017, tanvir2017, tominaga2018, valenti2017}, and from $\sim 10$ days after the merger, a broadband afterglow that appeared first at X-ray frequencies \citep{haggard2017, margutti2017, troya2017} and later at radio frequencies \citep{alexander2017, hallinan2017, kim2017}.
The early afterglow was due to a macronova \citep[a.k.a.~kilonova, e.g.,][]{barnes2016, tanaka2016, metzger2017} powered by the radioactive decay of $r$-process nucleosynthesis products \citep[e.g.][]{li1998, rosswog1998}.
The macronova faded rapidly over $\sim 5-10$ days.
The later broadband afterglow continued to rise over $\sim150$ days with optical observations by the Hubble Space Telescope confirming the single power-law from radio to X-ray frequencies at $\sim 90$ days \citep{lyman2018}, continued radio brightening $>100$ days \citep{resmi2018}, and a turn-over detected at radio and X-ray frequencies \citep{dobie2018, alexander2018, nynka2018}.

The late rising afterglow from GW170817 can be explained by a number of scenarios, including:
a structured jet, where the energy and velocity of the jet have some angular profile, or a jet-cocoon system, with the afterglow viewed off the central axis \citep[][etc]{davanzo2018, gill2018, lazzati2018, lamb2018, lyman2018, margutti2018, troya2018};
a choked or stalled jet that energises a cocoon \citep[e.g.][]{kasliwal2017, gottlieb2018, mooley2018, nakar2018};
fall-back accretion that re-brightens a jet afterglow \citep{li2018}; or
interaction of the dynamic tidal tails of the merger ejecta with the ambient medium \citep{hotoke2018}.
The prompt GRB 170817A does not rule out any of these scenarios \citep[e.g.,][]{bromberg2018, meng2018, zhang2018}. 

Due to GW selection effects, the most likely inclination for a GW detected merger is $\sim 31^\circ$ with a mean $\sim 38^\circ$ \citep{schutz2011, lambkobayashi2017}; therefore, there is a high probability  that future GW detections of neutron star mergers will have similar late rising afterglows.
It is difficult to distinguish between these afterglow scenarios on the basis of pre-peak signatures.  Polarization and resolved radio images at late times can provide information to help differentiate the choked-jet cocoon and jet models \citep{gill2018, nakar2018a}.
Here we show that the steepness of the decline post-peak can be used to differentiate an afterglow that is driven by an initially ultra-relativistic jet from a more mildly relativistic outflow, such as the cocoon or tidal tails.

In \S \ref{method} we describe our method for estimating the dynamics of a relativistic blast wave that becomes Newtonian, and describe how we estimate the synchrotron flux from the decelerating blast wave.
In \S \ref{results} we show the resultant light curves for the afterglows from the various outflow models, jet or cocoon, and in \S \ref{discussion} we discuss the results in detail and include the application to GW170817.
We give our concluding remarks in \S \ref{conclusions}.

\section{Method}
\label{method}

A rapidly expanding blast wave that propagates through an ambient medium will sweep up material and decelerate.
The ultra-relativistic dynamics of the decelerating blast wave can be explained by adiabatic and/or radiative expansion \citep{blandford1976, katzpiran1997, chiangdermer1999, dai1999, huang1999, piran1999, vanparadijs2000, bianco2004, bianco2005a, bianco2005b, peer2012, nava2013}.
For the dynamical parameters evolution we follow { the analytic solution of} \cite{peer2012} who considered, for the first time, both the energy density and the pressure of the shocked material.
{ These analytic solutions provide a good approximation for the expected behaviour of a blastwave, however, to accurately determine the energy and other parameters from the afterglow emission of a well sampled event the results of computationally more expensive full hydrodynamic simulations should be utilised \citep[see][etc]{kobayashi1999, granot2002, vaneerten2010, vaneerten2010a, vaneerten2012a, decolle2012, decolle2012a}.}
As we are concerned with the late-time evolution we will consider only the adiabatic case. 

\subsection{Dynamics}

Consider a blast wave of cold material with a kinetic energy $E = \Gamma_0 M_0 c^2$, where $c$ is the speed of light, $\Gamma_0$ the coasting phase bulk Lorentz factor, and $M_0$ the ejected mass. The change in Lorentz factor ${\rm d}\Gamma$ for an adiabatically expanding blast wave with swept-up mass ${\rm d}m$, in logarithmic bins, is \citep{peer2012}
\begin{equation}
\frac{{\rm d}\Gamma}{{\rm d}\log_{10} (m)} = -\frac{m\log{(10)} \left[\hat\gamma(\Gamma^2-1)-(\hat\gamma-1)\Gamma\beta^2\right]}{M_0+m\left[2\hat\gamma\Gamma-(\hat\gamma-1)(1+\Gamma^{-2})\right]},
\label{dyn}
\end{equation}
where $m$ is the swept up mass, $\beta = (1-\Gamma^{-2})^{1/2}=v/c$ is the normalized velocity, and $\hat\gamma$ is the adiabatic index
\footnote{The adiabatic index is found by \cite{peer2012} to be $\hat\gamma \simeq (5-1.21937z+0.18203z^2-0.96583z^3+2.32513z^4-2.39332z^5+1.07136z^6)/3$ where $z \equiv T/(0.24+T)$ and $T$ is the normalized temperature, $T \simeq (\Gamma\beta/3)([\Gamma\beta+1.07(\Gamma\beta)^2]/[1+\Gamma\beta+1.07(\Gamma\beta)^2])$.}.

\subsection{Synchrotron Emission}

The decelerating blast wave will accelerate electrons to relativistic velocities and enhance the downstream magnetic field.
These relativistic electrons will emit synchrotron radiation producing an observable broadband afterglow to the explosive event.

We assume that the blast wave accelerates electrons to a single power-law distribution of Lorentz factors ${\rm d} N_e/{\rm d}\gamma_e \propto (\gamma_e-1)^{-p}$ with a minimum Lorentz factor of $\gamma_m \geq 1$.  Although in practice we can expect the electron distribution to include a significant thermal population \citep{warren2018}, the effect of these thermal electrons is important at early times, and we ignore their contribution here.
The minimum electron Lorentz factor, where $p>2$, is $\gamma_m = 1+(p-2)/(p-1) \varepsilon_e (\Gamma-1) m_p/m_e$, where $\varepsilon_e$ is the fraction of downstream thermal energy in accelerated electrons, $m_e$ is the mass of an electron, and $m_p$ is that of a proton \citep{huang2003}.
The magnetic field $B'$ in the blast wave is calculated as ${B'}^2/(8{\rm\pi}) = \varepsilon_{\rm B}e$, where $\varepsilon_{\rm B}$ is the fraction of the energy that goes into the magnetic field. The thermal energy density is $e = [(\hat\gamma\Gamma+1)/(\hat\gamma-1)](\Gamma-1)n_0m_pc^2$ \citep{blandford1976, saripiran1995}, where $n_0$ is the ambient particle number density.

The cooling frequency, characteristic synchrotron frequency, and the peak spectral power for a power law distribution of electrons are found following \cite{wijersgalama1999} as: 
$\nu'_c = (0.286) 3\gamma_c^2 q B'/(4{\rm\pi} m_e c)$ for the co-moving frame cooling frequency, $\nu_m' = 3 X_p \gamma_m^2 q B'/(4{\rm\pi} m_e c)$ for the co-moving characteristic frequency, and $ P'_{\nu_m'} = f_x \sqrt{3} q^3 B'/(m_ec^2)$ for the co-moving peak spectral power per radiating electron.
Here $q$ is the charge of an electron, $X_p$ and $f_x$ are the dimensionless spectral maximum and dimensionless peak flux respectively
\footnote{The dimensionless parameters $X_p$ and $f_x$ are determined by solving the isotropic synchrotron function for a power law distribution of electrons. The synchrotron function is a Bessel function of the second kind \citep{rybickilightman1979}, the isotropic synchrotron function assumes that the emitting electrons have an isotropic distribution of angles between the velocity vector and the magnetic field, which is assumed to be tangled. Solving the isotropic synchrotron function for a power law distribution of electron energies gives a dimensionless spectrum, with spectral maximum and the flux at the maximum as $X_p$ and $f_x$. See Figure 1 in \cite{wijersgalama1999} for these quantities as a function of the electron index $p$.}, 
and primed quantities are in the co-moving frame throughout.
The electron Lorentz factor for efficient cooling is $\gamma_c = 6 {\rm\pi} m_e c/(\sigma_{\rm T}\Gamma {B'}^2 t_{{\rm obs,} 0})$ where $\sigma_{\rm T}$ is the Thomson cross-section and $t_{{\rm obs,} 0}$ is the time for an observer at an inclination $\iota=0$ as measured from the GW coalescence time. 

The maximum synchrotron specific flux is then $F_{\nu,{\rm max}} = N P_{\nu_m}/(\Omega D_L^2)$, where $N$ is the number of emitting electrons, $\Omega$ is the solid angle of the emission, $D_L$ is the luminosity distance\footnote{Here we neglect the redshift of the source. Where $z \gg 0$ the power becomes $(1+z)P_\nu$ \citep{kumar2015}}, and $P_{\nu_m} = \delta P'_{\nu_m'}$ is the observed maximum specific power per electron where $\delta = [\Gamma(1-\beta\cos i)]^{-1}$ is the Doppler factor and $i$ the angle between the observer and the emitting point's direction of bulk motion.
We assume that the co-moving emission is isotropic $\Omega' = 4{\rm\pi}$, and as the solid angle transforms as $\Omega = \Omega'/\delta^2$ \citep{rybickilightman1979}, the maximum specific power per steradian is $P_{\nu_m}/\Omega = \delta^3 P'_{\nu'_m}/\Omega'$, and the maximum specific flux for an observer at a distance $D_L$ and an inclination $i$ is \citep{salmonson2003}\footnote{As the shell has some thickness, the surface area change with inclination can be neglected.}
\begin{equation}
F_{\nu,{\rm max}} = \frac{N P'_{\nu_m'} \delta^3}{4{\rm\pi} D_L^2}.
\label{fmax}
\end{equation}
As the specific intensity in the co-moving frame is $I'_{\nu'} = N P'_{\nu'}/(\Omega' A)$, where $A$ is the emitting area, then the factor $\delta^3$ is consistent with the Lorentz invariant quantity $I_\nu/\nu^3 = I'_{\nu'}/{\nu'}^3$ \citep{granot2002}.
We note that this flux transform is valid for a point source \citep{ioka2017}.

The flux $F_\nu$ at a given observation time $t_{\rm obs}$ follows \cite{spn1998} and the time for an observer at $i$ is determined using
\begin{equation}
{\rm d} t_{\rm obs} = \left[\frac{1}{\beta(R)}-\cos{i}\right]\frac{{\rm d} R}{c},
\label{time}
\end{equation}
where $R$ is the radial distance of the blast wave in the lab frame.

\subsection{The Jet}

An ultra-relativistic outflow defined by the solid angle $\Omega_0 = 2{\rm\pi}(1-\cos \theta_0)$, where $\theta_0$ is the initial jet half opening angle and assuming $\Omega_0$ does not change, will have radius $R = [3 N/(\Omega_0 n_0)]^{1/3}$, where $N = m/m_p$ is the number of electrons for the calculation of $F_{\nu,{\rm max}}$ using equation \ref{fmax} and $m$ is the total swept up mass.
As the blast wave decelerates and the velocity becomes Newtonian $\Gamma\rightarrow1$, the minimum Lorentz factor for the electron distribution will also approach unity.
Electrons that are no longer ultra-relativistic will cease to emit synchrotron radiation, however, electrons with Lorentz factors between a cut-off value $\gamma_{\rm syn}$ and the maximum $\gamma_{M} = \Gamma t_{\rm obs} q B'/(m_p c)$ \citep{gao2013} will continue to emit synchrotron radiation.
By considering the distribution of electrons with a lower limit set by $\gamma_{\rm syn}\sim 2-5$ for the emission of synchrotron radiation, the number of synchrotron emitting electrons can be found following \cite{huang2003}.

If the outflow expands sideways, then the solid angle of the jet will increase \citep{huang2000a, huang2000b, huang2007, granot2005, vaneerten2010a, granotpiran2012, vaneerten2012a, granot2014}.
If the sideways expansion is set by the sound speed \citep[e.g.][]{huang2000a, huang2000b}
\begin{equation}
c_s = \left[\frac{c^2\hat\gamma(\hat\gamma-1)(\Gamma-1)}{1+\hat\gamma(\Gamma-1)}\right]^{1/2},
\label{cs}
\end{equation}
then the increase in the opening angle due to expansion is $\theta_j \approx c_s/(\Gamma \beta c)$ for small $\theta_j$ and roughly constant $c_s/(\Gamma \beta)$ \citep{rhoads1999,huang2000a}

For the purposes of determining the late-time afterglow we assume that the jets are homogeneous within a given opening angle, see \S\ref{results} for a discussion of the jet structure contribution to the afterglow at late times.
For each jet we consider multiple segments, or emission regions, and sum the flux over the equal arrival time surface \citep[e.g.][]{lambkobayashi2017, resmi2018}.
This ensures that we do not underestimate the flux for an off-axis observer due to the point-like assumption made in calculating the flux.

\subsection{The Cocoon}

If the jet has a low energy or there is a high mass of merger ejecta ($\gtrsim 0.05 M_\odot$) enveloping the polar regions where the jets are expected to be launched, then the material that the jet must drill through may be sufficient to stall or choke the jet \citep{ moharana2017, gottlieb2018, mooley2018}.  In fact, a jet propagating through dynamically ejected material should naturally produce a mildly relativistic $\Gamma < 10$ cocoon \citep{ramirezruiz2002, bromberg2007, bromberg2011, lazzati2017a, murguiaberthier2017a, nakar2017, gottlieb2018} over a much wider angle than a typical short GRB jet \citep{kasliwal2017, nakar2018}.  We therefore assume that all jets are accompanied by a cocoon and the cocoon emission represents the low-energy and low-$\Gamma$ limit for any jet structure. 
We model the afterglow light curve from this cocoon with the same method as for an ultra-relativistic jet, solving the dynamical equations for a blast wave with the lower $\Gamma_0$ and energy per steradian of the freely expanding cocoon \citep{lazzati2017a}. 
All the cocoons considered expand laterally at the sound speed.
By including the cocoon with each jet we show that the cocoon accompanying a jet does not contribute to the afterglow at late times, where the energy in the cocoon is less than the jet energy.

\section{Late-time Afterglows}
\label{results}

We produce late-time afterglow light curves for ultra-relativistic jets with and without sideways expansion.
For the sideways expansion we consider expansion at the sound speed \citep{huang2000b, huang2007}.
As \cite{vaneerten2012a} found relativistic outflows remain non-spherical even when $\Gamma\sim 2-3$, we assume that the sound speed expansion is an upper limit.
Additionally, light curves are produced for mildly relativistic cocoons.
These cocoons should accompany all jetted outflows and can contribute to, or be included in, jet structure descriptions \citep{lazzati2017, murguiaberthier2017, lambkobayashi2017, xie2018}.
Finally, we consider the afterglow from a choked-jet cocoon system, where the jet has failed to penetrate the merger environment and an energised cocoon will expand with an energy equivalent to that of the failed jet.

In each case we consider the dominant component at late times:
for the jets, this is the jet core and so we can ignore any structure that may be present at wider angles \citep[see][etc]{rossi2002, granot2005a, lambkobayashi2017, xiao2017, kathirgamaraju2018}.  The cocoon is included as the extreme limit of any jet structure and demonstrates that even for very wide angled cocoons the late time afterglow is dominated by the core;
for the choked jet-cocoon, which may have a stratified radial velocity profile \citep[e.g.][]{kasliwal2017, mooley2018}, we consider the component that would give the peak of the afterglow and dominate the late-time emission.

For our fiducial parameters we use the values shown in Table \ref{tab1}.
The jets have a half-opening angle of $6^\circ$, consistent with the limits from both \cite{ghirlanda2016} and \cite{fong2015}.
The cocoon accompanying a successful jet has a half-opening angle of either $20^\circ$ or $45^\circ$ and a maximum Lorentz factor $\Gamma = 3$ \citep{lazzati2017a, lazzati2017, gottlieb2018}, and the cocoon from a choked jet has an angle $40^\circ$ and a Lorentz factor $\Gamma = 1.4$ at the peak energy for the kinetic energy distribution of the radial component $E(\beta\Gamma)\propto (\beta\Gamma)^{-k}$  \citep{nakar2018a}.

For simplicity we use the same microphysical parameters $\varepsilon_e$ and $\varepsilon_{\rm B}$, as well as the same electron index $p$ for both the jet and cocoon afterglow models.
However, within the same source these parameters could be different for the cocoon and the jet afterglow.

\begin{table}
\caption{Fiducial parameters (isotropic-equivalent energy, initial Lorentz factor, initial half-opening angle) for the jet, cocoon, and choked jet dynamic and afterglow models. We assume an ambient density $n_0 = 0.01$ cm$^{-3}$, the microphysical parameter relation $\varepsilon_e = \varepsilon_{\rm B}^{1/2} = 0.1$ \citep{medvedev2006}, electron distribution index $p = 2.2$, and a distance of 100 Mpc for all models.}
\label{tab1}
\centering
\begin{tabular}{c c c c}
Parameter & Jet & Cocoon & Choked Jet \\
\hline
$E_{\rm iso}$ (erg) & $10^{51}$ & $10^{49}$ & $10^{51}$ \\
$\Gamma_0$ & $100$ & $3$ & $1.4$ \\
$\theta_0$ (deg) & $6^\circ$ & $20^\circ$ \& $45^\circ$ & $40^\circ$ 
\end{tabular}
\end{table} 

\begin{figure}
\includegraphics[width=\columnwidth]{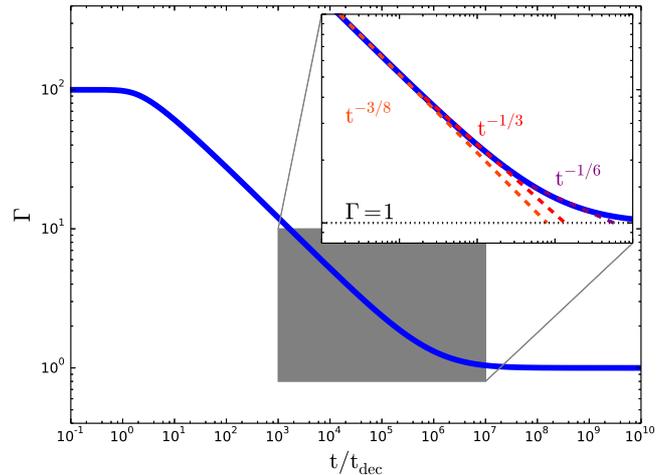}
\caption{The Lorentz factor with observer time in units of the deceleration time for a blast wave described by equation \ref{dyn}. The system has initial Lorentz factor $\Gamma_0 = 100$ and the jet fiducial parameters. The evolution of $\Gamma$ with observer time is shown for a jet that does not expand sideways. The zoomed section shows the deviation from the usually assumed $t^{-3/8}$ scaling when $\Gamma\lesssim5$.}
\label{fig1}
\end{figure}

The Lorentz factor $\Gamma$, the swept-up mass $m$, and the adiabatic index $\hat\gamma$ are determined for a spherical blast wave with the fiducial parameters by solving equation \ref{dyn} with a fourth order Runge-Kutta.
The Lorentz factor for a decelerating blast wave, with $\Gamma_0 =100$, is shown in Figure \ref{fig1}.
The $x$-axis is the observer time scaled by the deceleration time $t_{\rm dec}$, where $t_{\rm dec} \sim [3E_{\rm iso}/(256 {\rm\pi} n_0\Gamma_0^8m_pc^5)]^{1/3}$, and the zoomed section highlights how the commonly assumed $\Gamma \propto t^{-3/8}$ scaling for a decelerating blast wave changes at low Lorentz factors $\Gamma \lesssim 5$. Here the time is for an observer aligned with the jet's central axis.
This change in scaling will flatten the temporal index of the late decaying afterglow light curve as the blast wave approaches the Newtonian phase.

\begin{figure}
\includegraphics[width=\columnwidth]{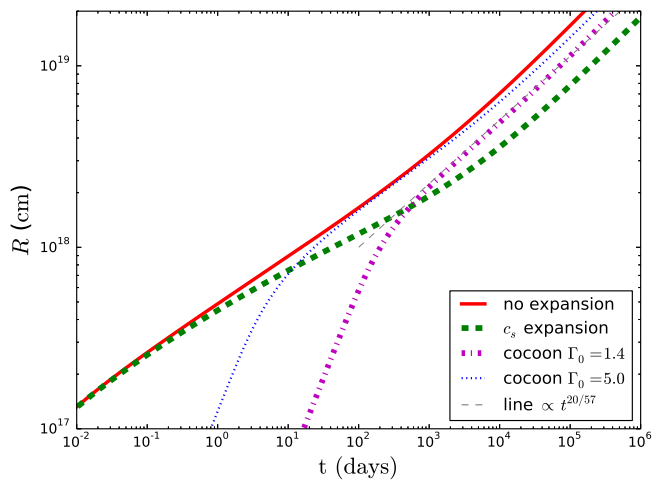}
\caption{The radius of the blast wave with time for an observer on the jet-axis. The two jet models are shown, with fiducial parameter values assumed. The red solid line is a jet without sideways expansion, the thick green dashed line is for a jet with the sound speed expansion. The choked-jet cocoon with fiducial parameters is shown as a pink dash-dotted line; this system has sound speed expansion and the initial deviation from $R\propto t^{2/5}$ is shown as a grey dashed line $\propto t^{20/57}$.  An additional cocoon with an initial $\Gamma_0=5$ is shown as a fine dotted line.}
\label{fig6}
\end{figure}

Where a jet expands sideways the radius of the shock front will differ from that of a non-expanding or a spherical blast wave.
Figure \ref{fig6} shows the radius of the blast wave with observer time for jets and a choked-jet cocoon with our fiducial parameters.
The effect of the jet expansion can be seen as a reduction in the blast wave radial velocity.
Before the jet expands, and during the ultra-relativistic phase, the radius scales with time as $R\propto t^{1/4}$.
The expansion of the jet reduces this scaling until the outflow becomes Newtonian, where the radius scales with time as $R\propto t^{2/5}$ \citep{huang1999}.
For an outflow with an initially low $\Gamma\lesssim2$, the scaling never follows the relativistic case and becomes Newtonian once the outflow sweeps up matter with a rest mass equivalent to the explosion energy.
The radius for an expanding low-$\Gamma$ outflow deviates from the expected $R\propto t^{2/5}$ and follows the slightly shallower $R\propto t^{20/57}$ shown by the grey dashed line\footnote{{ The exponent $20/57$ is a best fit line to the numerical integration.}}.
Where the initial Lorentz factor is $\Gamma_0\lesssim5$, we find the proportionality is between $t^{1/4}$ for $\Gamma_0>5$ and $t^{20/57}$ for $\Gamma_0\sim1.4$;
at $\Gamma_0\sim 2$ we find $\sim t^{1/3}$.

{ The emission from the blast wave is always dominated by the highest density region of the shell.
Although sideways expansion can be significant, the radial density profile of the shocked material was shown by \cite{kobayashi1999}, throughout the transistion from relativistic to Newtonian motion, to remain narrowly spiked at the shock front.
We have assumed that the emission is from a surface and ignore the shell thickness throughout the evolution.}

Using these parameters and the description for sideways expansion, the synchrotron flux at a given frequency from the blast wave for observers at an angle $0^\circ,~30^\circ,~60^\circ,~\&~90^\circ$ from the jet central axis is determined.
The jet with an inclination $\iota$, measured from the jet central axis to the observers line-of-sight, is assumed to be balanced by a counter-jet with an inclination $\iota+180^\circ$.
Similarly, a choked-jet cocoon will have a counter-cocoon system.
The emission from the counter-jet assumes symmetry of the system\footnote{The counter-jet, or the dynamical ejecta that the counter-jet must drill through, may not be identical to the near jet. We may see jet-cocoon and counter-choked-jet, choked-jet and counter-jet, or even single-sided systems, i.e., uni-polar outflows \citep[see ][]{tucker2017}.}; its high inclination yields a very small Doppler factor that significantly reduces the flux and elongates the observer timescale.
The observer time is additionally delayed by the extra distance traveled by the light due to the geometry of the system.
We assume the medium through which the light travels from the counter system to be optically thin.
The total flux for the afterglow is found by summing the cocoon, jet and counter-jet/counter-cocoon light curves to produce a total light curve at each time step.

\begin{figure*}
\includegraphics[width=\textwidth]{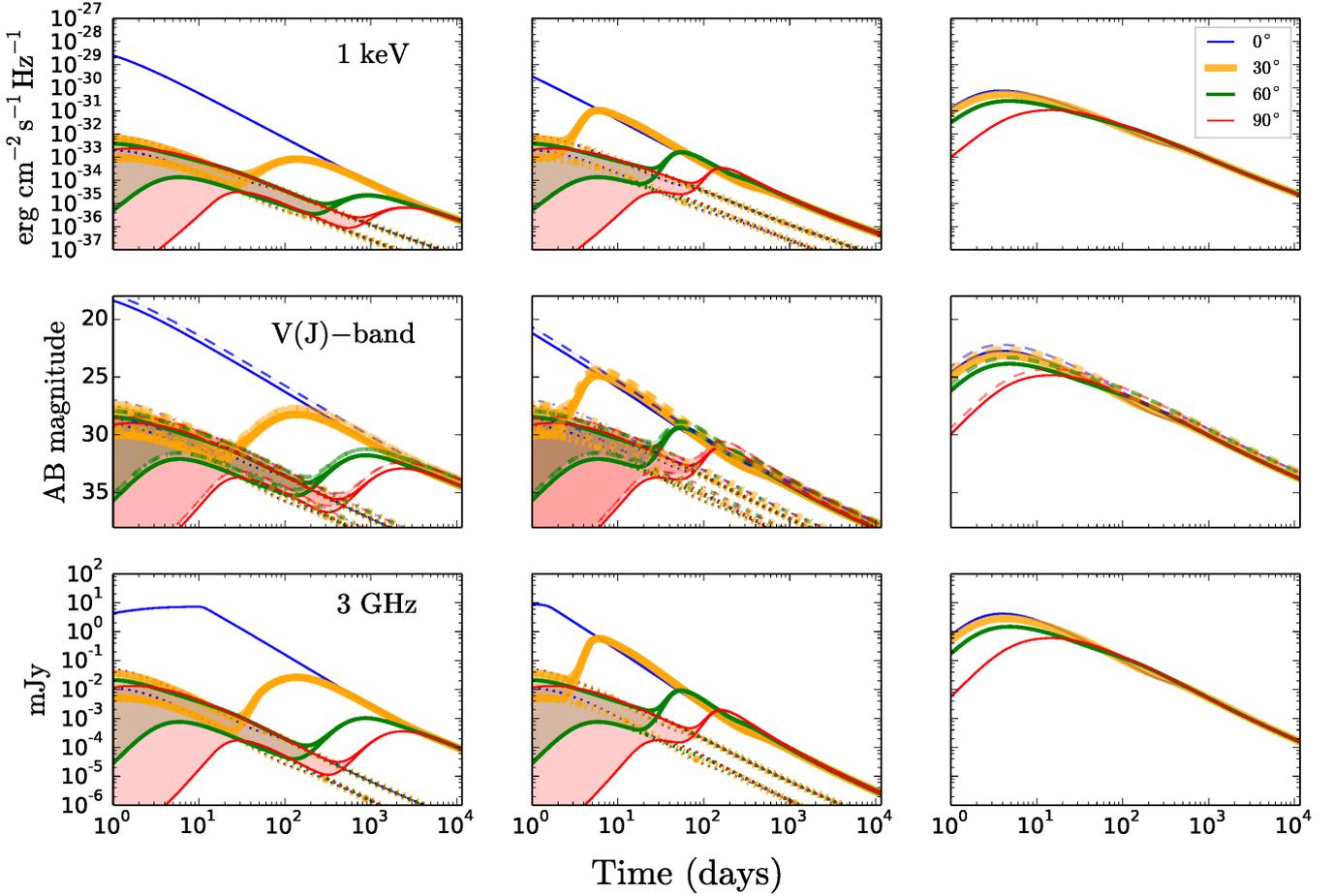}
\caption{Afterglow light curves from a decelerating relativistic outflow with fiducial parameters in Table \ref{tab1}, $n_0 = 0.01$ cm$^{-3}$, $\varepsilon_e = \varepsilon_{\rm B}^{1/2} = 0.1$, $p=2.2$, and a distance $D_L=100$ Mpc. The first two columns show the afterglow for a jet-cocoon system with the fiducial jet-cocoon parameters and jet expansion description: none, and sound speed. The shaded regions indicate the range for the jet-cocoon where the cocoon component has an angular width $20^\circ - 45^\circ$ and dominates the total flux; note the cocoon does not contribute at an inclination $\iota=0^\circ$. The third column shows the afterglow from a choked-jet cocoon. The rows show the flux at frequencies of 1 keV, V-band (and J-band as dashed and dot-dashed lines), and 3 GHz. The solid line in each panel is the total flux, the dotted line in the two jet-cocoon model afterglows shows the cocoon emission. Inclinations of $0^\circ,~30^\circ,~60^\circ,\&~90^\circ$ are colour an linewidth coded as thin blue, thick orange, medium green, and thin red respectively.}
\label{fig2}
\end{figure*}

The light curves at X-ray (1 keV), optical (V-band), IR (J-band), and radio (3 GHz) frequencies are shown for the jet cocoon systems (in the first two columns) and the choked jet system (in the last column) in Figure \ref{fig2}.  The X-ray afterglow is shown in the top row, the middle row shows both the V-band (solid line) and the J-band (dashed line), and the bottom panel shows the 3 GHz radio afterglow.
The two jet columns are respectively: no sideways expansion of the jet; and sideways expansion set at the sound speed.
The solid (and the dashed lines for the middle row) represent the total afterglow emission from both the cocoon and the jet.
The dotted lines (and dash-dotted lines in the middle row) in the first two columns indicate the cocoon emission only.
The limits of the shaded region indicate the jet-cocoon emission where the cocoon is contained initially within an opening angle $\theta_0 = 20^\circ$ or $45^\circ$.
The cocoon in a jet-cocoon system dominates the afterglow at early times for an observer at an inclination higher than the jet half-opening angle. 
The line colour and thickness indicates the observation angle: thin-blue is $0^\circ$; thick-orange is $30^\circ$; medium-green is $60^\circ$; and thin-red is $90^\circ$.
As we only consider the late-time afterglow we ignore any reverse-shock or self-absorption which is important at early times and at lower frequencies.

\begin{figure}
\includegraphics[width=\columnwidth]{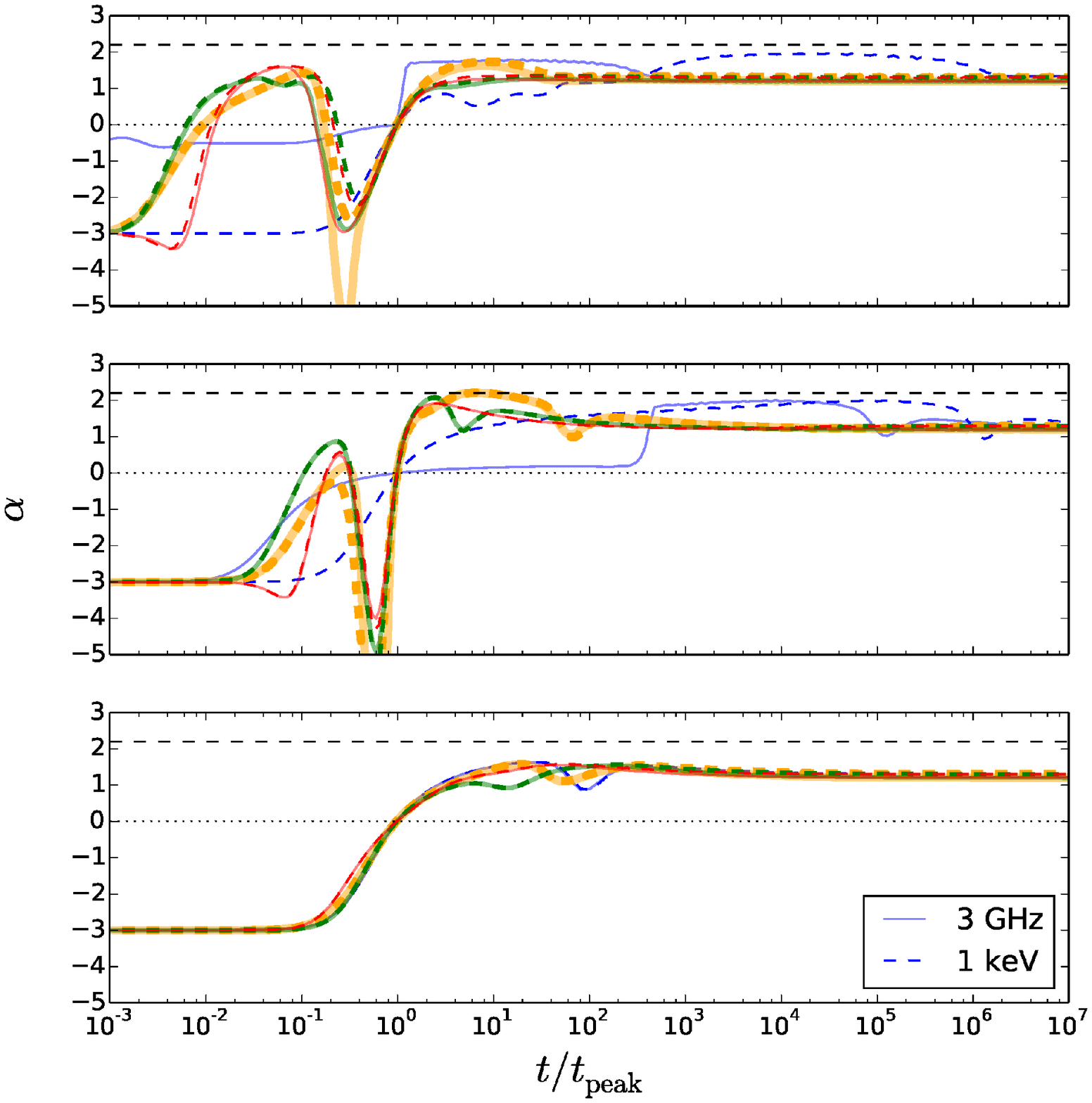}
\caption{The temporal index $\alpha$ for the afterglow as a function of the jet core afterglow peak time. The top panel shows the jet-cocoon afterglow index for a non-expanding jet, the second panel is the jet-cocoon afterglow index for a jet expanding at the sound speed, the third panel the temporal index for a choked-jet cocoon system. The index at 3 GHz is shown as a thick solid line, and at 1 keV with a dashed line. The line colour and width indicates the inclination to the observer's line-of-sight, $0^\circ,~30^\circ,~60^\circ,\&~90^\circ$ as in Figure \ref{fig2}. Horizontal black dotted line indicates an index of 0, and the black dashed line the fiducial electron index value $p = 2.2$.}
\label{fig3}
\end{figure}

The temporal index $\alpha$, where $F_\nu \propto t^{-\alpha}$, of the various afterglows is shown as a function of the jet core peak time in Figure \ref{fig3}.
Here the colour and linewidth scheme is the same as Figure \ref{fig2}.
The top two panels show the jet-cocoon systems with no expansion, and sound speed expansion.
The third panel shows the choked-jet cocoon system.
In each case the 3 GHz radio flux temporal index and the 1 keV X-ray flux index are shown as a solid line and a dashed line respectively.
The electron distribution index $p$ is shown as a dashed black line and a dotted black line indicates an index of 0, marking the transition from an increasing to a decaying light curve.

For jets, both with and without any sideways expansion, the temporal index of the post break decline (or post-peak decline for a moderately off-axis observer $\lesssim  60^\circ$) at its steepest is always larger than the equivalent for a wider angled choked-jet cocoon system.
At late times the jet in a jet-cocoon system, or the core in a structured jet, dominates the light curve.
Here structured jet refers to an angular dependence within the jet half-opening angle for the energy and Lorentz factor \citep[e.g.][]{rossi2002,granot2005a}.  For a structured jet the cocoon can be included as the low-energy limit of the structure \citep[e.g.][]{lazzati2017, lambkobayashi2017} although it is technically a distinct component situated beyond the jet half-opening angle.
In the case of the expanding jet, a flattening or a re-brightening of the light curve are seen due to the counter-jet.
A similar, although more modest effect, can be seen for the choked-jet cocoon system, where we assume that both jets are successfully stalled.

The transition from an ultra-relativistic outflow $\Gamma\gtrsim5$ to a Newtonian outflow $\Gamma\sim 1$ will affect the temporal index $\alpha$ of the decaying afterglow.
From Figure \ref{fig1}, where the outflow does not expand, we can see that $\Gamma\propto t^{-1/3}$ when $5\gtrsim\Gamma\gtrsim2$.
When the beaming angle of the emission becomes larger than the opening angle of the outflow, the edge of the jet will become visible resulting in the edge effect.
The edge effect will steepen the decline in the afterglow as $\Gamma^{-2}$.
For a relativistic jet, where $\Gamma \propto t^{-3/8}$, the temporal index due to the edge effect will steepen by $3/4$.
When $5>\Gamma\gtrsim2$ the edge effect results in the temporal index steepening by $\approx 2/3$.
However, when the Lorentz factor is  $\Gamma \lesssim 2$, the outflow will begin to transition to a Newtonian blast wave.
The temporal index in the Newtonian regime is $\alpha=3(5p-7)/10$ where $\nu_m<\nu<\nu_c$, and $\alpha=(3p-4)/2$ where $\nu_m<\nu_c<\nu$ \citep{huang2003, gao2013}.
For a low-$\Gamma$ outflow that experiences expansion during this Newtonian transition, the expansion slows the blast wave so that $R\propto t^{20/57}$ (see the grey-dashed line in Figure \ref{fig6}) where $\Gamma_0\sim 1.4$, or $\sim t^{1/3}$ for $\Gamma_0\lesssim 2$, resulting in a steepening of the expected temporal index by $\sim 0.15$ or $\sim 0.2$ respectively.

The jet without sideways expansion shows the temporal index behaviour expected from the jet-edge effect only, with the index steepened by $3/4$ while $\Gamma>5$.
For this system, the effect of the change in the $\Gamma$ vs $t$ scaling relation for $\Gamma \lesssim 5$, can be seen between $t\sim1000$ days and $\sim6000$ days where $\alpha$ goes from $\sim 1.65$ to $\sim 1.15$ for the flux at 3 GHz. 
For the expanding jet, the post jet break behaviour is initially dominated by the edge effect.
However, at times $\gtrsim10$ days for our parameters, the jet spreading slightly steepens the decline as the temporal evolution of the radius slows down.
For the choked-jet cocoon system, the decline steepens for a short period, the counter-cocoon contributes to the light curve from $\sim 1000$ days and at $\gtrsim 4000$ days the system starts to approach the Newtonian decline.

\section{Discussion}
\label{discussion}

We have used dynamical parameter evolution that traces the transition from a relativistic to a Newtonian blast wave to determine the late time behaviour of the afterglow light curves for jetted and wide angle cocoon-like emission expected to follow a gravitational-wave detected compact binary merger involving a neutron star.
The late time temporal index of the evolving counterpart can be used along with knowledge of the electron energy distribution index $p$, to identify the existence of an initially ultra-relativistic jet for observers at inclinations $\iota\gtrsim\theta_0+1/\Gamma_0$, where no bright GRB with a jet origin is expected
\footnote{For observers at $\iota>\theta_0+1/\Gamma_0$  a burst of gamma-rays similar to a short GRB may originate at the shock breakout from the merger ejecta, as may have happened with GRB 170817A \citep{granot2017}. Although a prompt GRB can be fully suppressed even for an observer at $\iota<\theta_0+1/\Gamma_0$ for a mildly relativistic jet $\Gamma_0\lesssim20$, such jets are expected to have afterglows similar to those from GRB producing jets \citep{cenko2013, lamb2016}.}.
Additionally, the late time behaviour can be used to put constraints on the degree of sideways expansion of the afterglow producing material.

For a jetted outflow, the rate of sideways expansion contributes to the timescale for the expected peak of the jet afterglow emission.
Timescales for jets with sideways expansion are shorter than those for outflows that have no expansion.
A jet with our sound speed expansion description will have a deceleration, or peak time for an off-axis observer, earlier by about an order of magnitude than for a jet without expansion.
The peak flux of the afterglow for an off-axis observer is brighter ($\sim 1-2$ magnitudes at optical frequencies for our parameters) than that for an identical jet but with less or no sideways expansion.

The post-jet-break, or post-peak for an off-axis observer, temporal decay can indicate the degree of sideways expansion.
For a jet without expansion the post-break or steepest decline index is determined by the jet edge effect.
The afterglow decline is shallower for observers at higher inclinations, and for $\iota\gtrsim 60^\circ$ the afterglow decline approaches the Newtonian solution with little evidence of a jet.
For a jet that expands sideways, at wider inclinations the counter-jet will indicate the jet-like nature of the afterglow before the transition to the Newtonian phase.
Although we have not considered very rapid expansion, as it is not seen in hydrodynamic simulations \citep{vaneerten2012a}, for a jet that expands more rapidly than the sound speed case the post-peak decline for an off-axis observer will be at an index $\alpha>p$ \citep{vaneerten2010a}.
After the steepest decay phase the afterglow temporal index will decrease from the rapid decay phase until it approaches the Newtonian solution, interrupted only by the counter-jet.

The jet-cocoon light curves presented in Figure \ref{fig2}, first two columns, assume a homogeneous jet surrounded by a cocoon.
Within the jet half-opening angle, the jet is expected to have a degree of angular structure and the transition from the jet to the cocoon will be `softer' due to shear forces and mixing than the sharp edge assumed here.
Additionally, the cocoon can be expected to have some intrinsic angular structure.
This angular structure within the jet and the cocoon system will result in an afterglow light curve with the expected features of a structured jet profile discussed in relation to GW detected counterparts by \cite{lambkobayashi2017}, \cite{lazzati2017a}, \cite{jin2018}, \cite{kathirgamaraju2018}, \cite{xie2018} and others.
The shape of the pre-peak light curve for an off-the-jet-axis observer can give an indication of the jet-cocoon angular structure \citep{lambkobayashi2017}.

Throughout we assume that the core in the jet structure description is the fastest and most energetic component.  Alternatively the jet structure could be dominated by other regions within the opening angle such as with anisotropic jets \citep[e.g.][]{yamazaki2004,ioka2005} or an energetic second component \citep[e.g.][]{barkov2011}.  For such jets the most energetic component would still dominate the late-time afterglow.
Variability or re-brightening of the late afterglow could indicate either multiple `mini-jets' or an energetic second-component.

We have assumed plausible parameters for our jets, cocoons, and choked-jet cocoons.
The choice of parameter values alter the timescale and peak flux of the light curves shown in Figure \ref{fig2}.
The timescales for deceleration and the transition to the Newtonian regime depend on the environment ambient number density as $t\propto R\propto n_0^{-1/3}$.
As $F_{\nu, \rm{max}} \propto E n_0^{1/2} \varepsilon_{\rm B}^{1/2} D_L^2$, the brightness of the afterglow is most sensitive to these parameters.  
We have only considered jets accompanied by cocoons with less energy than the jet;
if a population of choked-jet cocoons exists \citep{moharana2017} then there must be a transitional parameter region where the cocoon has a higher energy than a successful jet.
In such a case we may see the cocoon at late times even for an on-axis observer, for an off-axis observer this energetic cocoon would dominate the decline at late times and hide the successful but weak GRB producing jet.

The observation of an afterglow emission from a jet at the transition to the Newtonian regime will require very sensitive telescopes at all wavelengths, such as Chandra at 1 keV, the Hubble Space Telescope and the James Webb Space Telescope at optical and infrared wavelengths, and the Square Kilometre Array at radio frequencies.
For a source at $100$ Mpc and our parameters, the required sensitivity is $\sim 10^{-35}$ erg cm$^{-2}$ s$^{-1}$ Hz$^{-1}$ at 1 keV, AB magnitude $>32$ at optical and infrared wavelengths, and $\sim 10^{-1}$ $\mu$Jy at 3 GHz.

\subsection{Application to GW170817}

After the rapidly fading macronova associated with GW170817 had vanished from view \citep{lyman2018}, the late time afterglow was observed to increase gradually over a period $\sim 10-150$ days \citep[e.g.][]{alexander2018}.
This enigmatic afterglow can be fit by various structured jet/jet-cocoon systems or a choked-jet cocoon with a radial velocity profile.
To fit the early afterglow, all the structured jet/jet-cocoon models have a jet core that is more energetic than the cocoon and so will always dominate the late time afterglow.
%Additionally the high velocity dynamic tidal tails of the merger ejecta can offer a tentative fit to this long lasting afterglow \citep[see][]{hotoke2018}.
%Here we consider only the jet and the choked-jet afterglow prediction.  We expect the afterglow from a tidal tail to behave similarly to that from a wide-angle low-$\Gamma$ outflow such as the choked-jet, potentially making it challenging to distinguish dynamic tidal tail afterglows and choked-jet cocoons.

\begin{figure}
\includegraphics[width=\columnwidth]{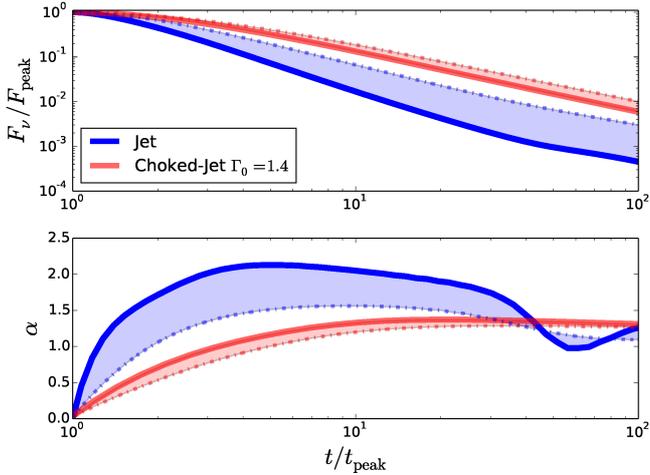}
\caption{The top panel shows the post-peak afterglow flux at 3 GHz for a jet viewed at $\iota = 29.5^\circ$ as a blue shaded region and the post-peak afterglow flux from a choked-jet cocoon viewed at $\iota = 20^\circ$ as a red shaded region, where the cocoon has an initial opening angle $\sim 40^\circ$.
The shaded region covers the range of sideways expansion prescriptions where a dash-dotted line indicates no expansion and a thick solid line sound speed expansion.
The electron index is $p=2.11$ for the jet \citep{lyman2018} and $p=2.2$ for the choked-jet cocoon \citep{mooley2018}.
The flux and time are normalised to the peak value. 
The bottom panel shows the temporal index post-peak for each system where the shading and line style are the same as the top panel.}
\label{fig5}
\end{figure}

In Figure \ref{fig5} we show the normalised flux at a frequency of 3 GHz and normalised time post-peak for the emission from a jet with the parameters from \cite{lyman2018}, $p=2.11$ and $\iota=29.5^\circ$, and a choked-jet cocoon with parameters from \cite{mooley2018} and a $\Gamma_0=1.4$, $p=2.2$ and viewed within the cocoon half-opening angle.
The shaded regions indicate the parameter space between the expansion descriptions for the jet and cocoon.
As the temporal index $\alpha$ during the declining phase depends only on $\Gamma$ and $p$ (the jet models require an inclination outside of the jet core angle) we can ignore the difference in the model parameters by presenting the flux and time as a fraction of the peak values.
If both systems behaved in a similar fashion post peak, then we would expect the choked-jet cocoon to have the steepest decline index as we assume a $p$ value that is slightly larger than that for the jet system.
However, it is clear from Figure \ref{fig5} that if the afterglow is due to an off-axis jet, where the core is now dominating the emission, we will see an afterglow that declines steeper than that achieved by the choked-jet cocoon system, $\alpha \gtrsim 1.5$.
Conversely, if the declining afterglow never becomes steeper than $\alpha \lesssim 1.35$, then a choked-jet cocoon peaking with $\Gamma_0\sim1.4$ is most likely responsible for the entire afterglow from $\sim 10$ days post merger.

\cite{dobie2018} reported that the afterglow luminosity peaked in the radio at $\sim 150$ days.
The subsequent decline in the afterglow has been confirmed at X-ray frequencies with observations at $\sim 260$ days post merger showing a decline since $\sim 160$ days \citep{alexander2018,nynka2018}.
Extrapolating a power-law decay from just these two data points, the decline in X-ray flux indicates that between these observations, the afterglow has declined with an index of $\alpha\sim 1.3$, suggesting the start of rapid expansion.
This decline is measured over a period $\sim1.6~\times$ `peak time', assuming a peak at $\sim 150-160$ days.
This is consistent with the jet afterglow model where the index at this time should be $0.7\lesssim\alpha\lesssim1.4$.

If the decline follows the off-axis jet afterglow model, the flux will decay by an order of magnitude over a duration $\sim4~\times$ `peak time', or $\lesssim 2$ years.  This would bring the X-ray flux close to the {\it Chandra} detection limit based on Figure \ref{fig5} (although this figure was computed for a radio afterglow, the decay will be similar at X-ray frequencies, especially for the cocoon where the cooling frequency $\nu_c$ is above the X-ray band even at late times).
For the Very Large Array, the sensitivity is $\sim 10$ $\mu$Jy, approximately an order of magnitude fainter than the peak flux, giving the same time scale as for the X-ray flux.
There should be $>3$ observations by each of these telescopes over the respective time scale to get a reliable constraint on the temporal index $\alpha$ of the decline.
At X-ray and radio frequencies a temporal index $\alpha\gtrsim1$ will strongly favour an initially ultra-relativistic and off-axis jet driving the late time afterglow.  
A wide-angle outflow will be above the sensitivity limits for much longer than the jet afterglow.  

Considering energy and ambient number density constraints, analytic estimates favour $\Gamma_0\lesssim2$ for a wide-angle outflow that peaks at $\sim150$ days.
While we focus on two alternative models -- a successful jet with an ultra-relativistic core $\Gamma_0 \gtrsim 100$ or a choked jet transferring energy to a wide-angle, mildly relativistic cocoon where the peak time emitting component has $\Gamma_0 \lesssim 2$ -- intermediate models have been proposed in the literature.  
For example, \cite{xie2018} suggest a wide-angle cocoon with $\Gamma_0 \sim 10$. 
Because the light-curve decay at late times is driven by a fast moving component of the ejecta outflow, such cocoon models will behave similarly to our non-expanding relativistic jets, although the break to the steepest decline will be later than the peak time.

If the current afterglow $\lesssim300$ days is purely an energetic cocoon component then a contribution from a narrow and highly energetic jet may appear at $\sim 1000$ days \citep{barkov2018}; in this scenario a cocoon peaks and then declines and the jet afterglow will cause a re-brightening of the light curve at very late times.
The two episodes will have declining light curves that follow the temporal indices described here.

\section{Conclusions}
\label{conclusions}

We have shown that the late-time, post-peak decline in the afterglow of a GW-detected merger system can be used to differentiate between outflow geometries. 
With the aid of constraints on the system's inclination from the gravitational-wave signal, the steepest decay index post peak for a mildly inclined event will differentiate between any of the jet models and a choked-jet cocoon.

For an afterglow that is dominated by the decelerating core of an initially ultra-relativistic jet the late time decline will show evidence of an edge effect, or expansion, by exhibiting a temporal index steeper by $>3/4$ than the well-known $(3p-3)/4$ or $(3p-2)/4$ for $\nu_m<\nu$ or $\nu_c<\nu$ emission. 
Mildly off-axis $\iota \lesssim 60^\circ$ observers may see a very steep decline, depending on the details of the jet expansion.

For the choked-jet cocoon, due to the low-$\Gamma$ value for such systems, the temporal index is determined by the relativistic to Newtonian transition.
Where the cocoon expands rapidly we see a slight steepening of the Newtonian regime temporal index, as $F_\nu \propto R^3 t^{(9-15p)/10}$ \citep{dai1999}, and $R\propto t^k$ where $k\sim 1/3$ for $\Gamma_0\lesssim 2$ and $\sim 20/57$ for $\Gamma_0 = 1.4$ instead of the expected $R \propto t^{2/5}$.

The post-peak afterglow will decline with a temporal index $F_\nu \propto t^{-\alpha}$ for:
\begin{itemize}
\item Jet; $\alpha>3p/4$ for $\nu_m<\nu$ and $\alpha>(3p+1)/4$ for $\nu_m<\nu_c<\nu$;
\item Choked-jet cocoon where $\Gamma_0 \lesssim 2$; $\alpha<(15p-19)/10$ for $\nu_m<\nu$ and $\alpha<(15p-18)/10$ for $\nu_m<\nu_c<\nu$;
\item Choked-jet cocoon where $\Gamma_0 = 1.4$; $\alpha<(855p-1113)/570$ for $\nu_m<\nu$ and $\alpha<(285p-352)/190$ for $\nu_m<\nu_c<\nu$.
\end{itemize}

Applying this technique to the afterglow of GW170817, we find that if the declining afterglow is observed to have a temporal index $\alpha\gtrsim 1.5$, at its steepest, then only the core of a decelerating jet can achieve this.
In contrast, if the decline is always $\alpha\lesssim1.4$, then a choked-jet cocoon or dynamic tidal tails are most likely responsible for the afterglow.
Observations at radio and X-ray frequencies from peak flux until $\sim 10 \times$ the peak flux time are crucial to distinguish between the jet and the choked-jet afterglow.  Based on just two X-ray data points at 160 and 260 days, the rate of the afterglow decline is consistent with an off-axis jet afterglow.

Using the most sensitive telescopes, and relying on nearby mergers $\lesssim 40$ Mpc, the point at which a system becomes Newtonian can be probed.
A change, or brightening, of the afterglow decline can indicate the contribution from a counter-jet, the degree of this contribution can constrain the sideways expansion of an initially ultra-relativistic jet. 

\section*{Acknowledgements}
We thank the referee for their constructive comments.
GPL thanks Shiho Kobayashi for useful discussions.
GPL is supported by a Science Technology and Facilities Council Grant (STFC) ST/N000757/1.  
IM acknowledges partial support from the STFC. 
RL acknowledges support from the grant EMR/2016/007127 from Dept. of Science and Technology, India.

\bibliography{main.bib}{}
\bibliographystyle{mnras}
\bsp
\label{lastpage}
\end{document}